\documentclass{aa}
\usepackage{graphicx}
\usepackage{natbib}
\newcommand{\phn}{\phantom{0}}

\begin{document}

\title{3-D Kinematics of the HH~110 jet
\thanks{Partially based on observations made with the 2.5~m Isaac Newton 
and the Nordic Optical Telescopes 
operated at the Observatorio del Roque 
de los Muchachos of the Instituto de Astrof\' \i sica de Canarias. } 
}

\author{
Rosario L\'opez \inst{1}
\and Robert Estalella \inst{1}
\and Alejandro C. Raga \inst{2}
\and Angels Riera \inst{3,1}
\and Bo Reipurth \inst{4}
\and 
Steve R. Heathcote\inst{5}
}

\offprints{R. L\'opez}

\institute{
Departament d'Astronomia i Meteorologia, Universitat de Barcelona,
Av.\ Diagonal 647, E-08028 Barcelona, Spain.\\
\email{rosario@am.ub.es, robert.estalella@am.ub.es}
\and
Instituto de Ciencias Nucleares, Universidad Nacional Aut\'onoma de
M\'exico, Apartado Postal 70-543, 04510 M\'exico D.F., M\'exico.\\
\email{raga@nuclecu.unam.mx}
\and
Departament de F\'{\i}sica i Enginyeria Nuclear,
Universitat Polit\`ecnica de Catalunya, Av. V\' {\i}ctor Balaguer s/n, E-08800
 Vilanova i la Geltr\'u, Spain. \\
\email{angels.riera@upc.es}
\and
Institute for Astronomy, University of Hawaii, 2680 Woodlawn Drive, Honolulu, HI
96822, USA. \\
\email{reipurth@ifa.hawaii.edu}
\and
Southern Astrophysical Research Telescope, Casilla 603, La Serena, Chile.\\
\email{sheathcote@ctio.noao.edu}
}

%\date{Received ..}

\abstract{
We present new results on the kinematics of the jet HH~110. New proper motion
measurements have been calculated from [SII] CCD images obtained with a time
baseline of nearly fifteen years. HH~110 proper motions show a strong 
asymmetry with respect to the outflow axis, with a general trend of pointing 
towards
the west of the axis direction. Spatial velocities have been obtained by
combining the proper motions and  radial velocities from Fabry-P\'erot data.
Velocities decrease by a factor $\sim3$ over a distance of $\sim10^{18}$~cm,
much shorter than the distances expected for the braking caused by the
jet/environment interaction. Our results show evidence of an anomalously strong
interaction between the outflow and the surrounding environment, and are
compatible with the scenario in which HH~110 emerges from a deflection in a
jet/cloud collision.
}

\maketitle

\section{Introduction}

The HH~110 jet (which is found in the Orion B cloud complex) was
discovered by Reipurth \& Olberg \cite{rei91}. 
This jet presents a morphology in H$\alpha$ and [SII] images that is 
quite different from the
morphologies of other well known stellar jets. While the best studied jets
(e.g., HH~111 and HH~34) show chains of aligned knots with more or less
organized, arc-like shapes, the HH~110 jet has a more chaotic structure:
it starts in a well collimated chain of knots and then widens in a cone of
opening angle $\sim10^{\circ}$, with a rather chaotic knot structure
and appreciable wiggles along the length of the jet. The observed
structure is  reminiscent of a turbulent outflow.

In contrast to the optical images, the HH~110 jet is nearly straight in
near infrared H$_2$ 2.12~$\mu$m images. Furthermore, the H$_2$ emission appears
shifted westward relative to the optical emission (Noriega-Crespo et al.\
\citealp{nor96}).

Another special feature of the HH~110 jet is the lack of detection of any
stellar source (suitable for powering the jet) along the outflow axis. The
morphology of HH~110 first suggested that the driving jet source would be
embedded in a dark lane located to the north of the apex of the outflow.
However, searches at optical, near infrared and radio continuum
wavelengths have failed to detect the HH~110 driving source.

Reipurth, Raga \& Heathcote \cite{rei96} report the discovery of another
fainter jet, HH~270, located $3'$ northeast of HH~110.  An embedded 
near-infrared source very close to IRAS~05489+0256 was also found along the
HH~270 flow axis. They suggest that HH~110 is the result of the
deflection of the HH~270 jet (which travels in an E-W direction) through a
collision with a dense molecular cloud core. In this scenario, HH~270
strikes the molecular core and is deflected, giving rise to the HH~110
jet. The H$_2$ emission would then be a tracer of the region where the atomic
jet and the molecular cloud core interact (Noriega-Crespo et al.\ \citealp{nor96}).

Because of the interesting dynamics resulting from the interaction between
a radiative jet and a dense obstacle, a number of theoretical papers on
the HH~270/HH~110 system have been written. Raga \& Cant\'o \cite{rag95}
explored the initial stages of a jet/cloud collision with both
analytic models and time-dependent, 2D numerical simulations. The
final, steady state in which the jet has bored a hole through the
(stratified) cloud was studied analytically and numerically
by Cant\'o \& Raga \cite{can96} and Raga \& Cant\'o \cite{rag96}.
The initial state of a jet/cloud interaction (which we now appear
to be seeing in the HH~270/HH~110 system) was re-explored, now
with 3D numerical simulations, by de Gouveia dal Pino \cite{gou99}.
Hurka, Schmid-Burgk \& Hardee \cite{hur99} carried out 3D numerical
simulations of a jet interacting with a magnetized cloud.
The most recent paper on models of a radiative jet/dense cloud interaction
is the one of Raga et al.\ \cite{rag02}, who carried out 3D simulations
with a simplified atomic/molecular network in order to obtain predictions
of atomic and molecular emission line maps for carrying
out direct comparisons with the HH~270/HH~110 system.

The HH~110 jet seems to be appropriate for searching the observational
signatures of entrainment and turbulence through spatially resolved
kinematic studies. Recently, Riera et al.\ \cite{{rie03a},{rie03b}} 
presented
detailed radial velocity studies obtained with multi-long-slit
spectroscopy and Fabry-P\'erot interferometry, and compared the
observations with simple, parametrized models of a turbulent jet ( 
Cant\'o, Raga \& Riera \citealp{can03}).

In order to complement these radial velocity studies, we now present new
proper motion measurements of HH~110. These proper motions have been
determined from deep [SII] CCD images that were obtained with a time
baseline of 5424 days (i.e., nearly fifteen years), and show the
kinematical properties of a large number of knots. The only previous
proper motion determinations of HH~110 are those of Reipurth et al.\
\cite{rei96}, who study images with a shorter time baseline ($\sim6$~yr) and
are not able to reach a level of accuracy and detail comparable to that
 of our present measurements.
 
\begin{table*}
\caption[ ]{Log of the Data}
\begin{tabular}{lrrrr}  
\hline\hline
Epoch & Telescope & Pixel scale & Exp. Time & Reference \\
      &           &(arcsec)     & (s)        &          \\
\hline      
1987 Dec 18 & 1.5~m La Silla    &0.35 & 3600 & 1\\
1988 Mar  5 & 3.6~m ESO         &0.35 & 1800 & 1\\
1993 Dec 16 & 2.5~m INT         &0.55 & 9000 & 2\\
1994 Jan 15 & 3.5~m NTT         &0.35 & 1200 & 1\\
2002 Oct 24 & 2.6~m NOT         &0.19 & 9000 & 3\\
\hline
\end{tabular}
\begin{list}{}{}
\item[] References:
(1) Reipurth et al.\ \cite{rei96}; (2) Riera et al.\ \cite{rie03a};\\
(3) obtained by G. G\'omez through the NOT Service Time facility.
\end{list}
\label{t1}
\end{table*}

The paper is organized as follows. The observations are described in Sect. 2. 
In Sect. 3 we present the procedures to measure the proper motion velocities (Sect. 3.2)
and discuss the obtained results (Sect. 3.1). The radial and total velocities 
are
discussed in Sect. 4. Finally, the conclusions are given in Sect. 5.

\section{Observations}

In order to determine proper motions of HH 110 jet, a set of five CCD
images has been used, yielding a time baseline of 5424 days (nearly
fifteen years). Details of the epochs, telescopes, spatial resolution and
exposure times are given in Table~\ref{t1}. All the images were obtained
through [SII] narrow-band filters, which included the [SII]
$\lambda=6717,6731$ \AA\ emission lines. 
It is not possible to perform similar proper motion determinations from the 
H$\alpha$ emission, as we did not obtain images through an H$\alpha$ narrow-band  
filter during the observing runs carried out in the 1993 and 2002 epochs.

The 1987 image was obtained at the Danish 1.5~m Telescope at La Silla. The
1988 image was obtained at the ESO 3.6~m Telescope. The 1994 image was
obtained with EMMI at the ESO 3.5~m New Technology Telescope (NTT).
Details on the acquisition and treatment of all these data can be found in
the papers of Reipurth \& Olberg \cite{rei91} and Reipurth et al.\ \cite{rei96}.

The 1993 image was obtained at the prime focus of the 2.5-m Isaac Newton
Telescope (INT) of the Observatorio del Roque de los Muchachos (ORM, La
Palma, Spain). The detector used was a CCD with a coated EEV chip. A
filter of central wavelength $\lambda=6730$ \AA\ and bandpass
$\Delta\lambda=48$ \AA\ was used.

The 2002 image was obtained with the 2.6-m Nordic Optical Telescope (NOT)
of the ORM. The Andalucia Faint Object Spectrograph and Camera (ALFOSC) was used,
with a Ford-Loral CCD, and a filter of central wavelength
$\lambda=6724$ \AA\ and bandpass $\Delta\lambda=50$ \AA\ . Both
the 1993 and the 2002 images were obtained by combining five frames of
1800~s exposure each.  For these two epochs, the individual exposures were
reduced (bias-subtracted and flat-fielded) using the same procedure,
through the standard tasks of the IRAF \footnote{IRAF is distributed by
the National Optical Astronomy Observatories, which are operated by the
Association of Universities for Reseach in Astronomy, Inc., under
cooperative with the National Science Foundation} reduction package. The
five individual frames were recentered using the position of several field
stars, in order to correct for misalignments among the individual
frames. Then the frames were median-averaged using the IMCOMBINE task of
IRAF (with appropriate options to remove cosmic ray events). We obtained one
deep [SII] image with a total integration time of 2.5~h for each of these
two epochs (i.e., 1993 and 2002). These final images are not
flux-calibrated.

The central radial velocities of the main knots of HH~110 were obtained from
H$\alpha$ Fabry-P\'erot data (see Riera et al.\ \citealp{rie03b} for 
more details of
the observations and the data reduction).

\section{Proper motion measurements}

\begin{figure}
\resizebox{0.55\hsize}{!}{\includegraphics{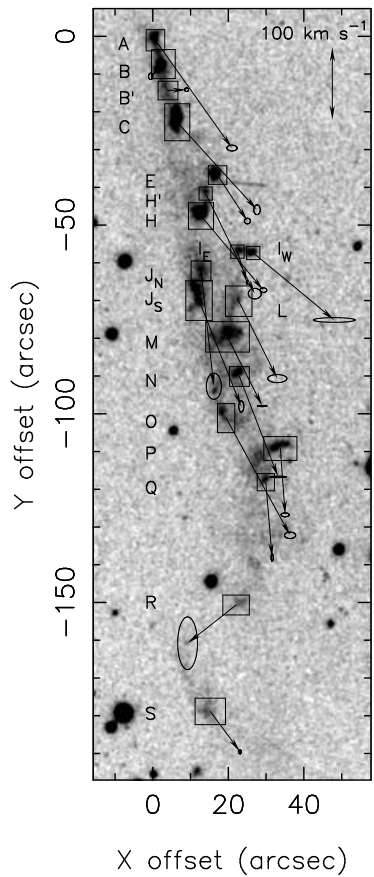}}
\caption{[SII] 6717+6731 image of HH~110 from the October 2002 NOT image.  
Arrows indicate the proper motion velocity of each knot. Ellipses at the
end of each arrow indicate the uncertainty in the components of the
velocity vector. The scale of the arrows is indicated by the double headed
arrow at the top right corner of the map, and corresponds to a velocity of
100~km~s$^{-1}$.  Boxes indicate the regions used for the proper motion
calculations. Offsets are measured from the peak position of knot A. North
is up and East is to the left. 
\label{fig1}} 
\end{figure}

\subsection{Method and results}

Proper motions of the main HH 110 knots have been determined
from the CCD images listed in Table~\ref{t1}. First, the five images were
converted onto a common reference system and rebinned to the same pixel
scale. We used the position of fifteen common field stars, in order to
register the images. These reference stars are well distributed around the
HH 110 jet. The GEOMAP and GEOTRAN tasks of IRAF were applied to perform a
linear transformation with six free parameters that take into account
relative translation, rotation and magnifications between the frames.  
The final, transformed frames have a pixel size of $0\rlap.''35$.

\begin{figure*}
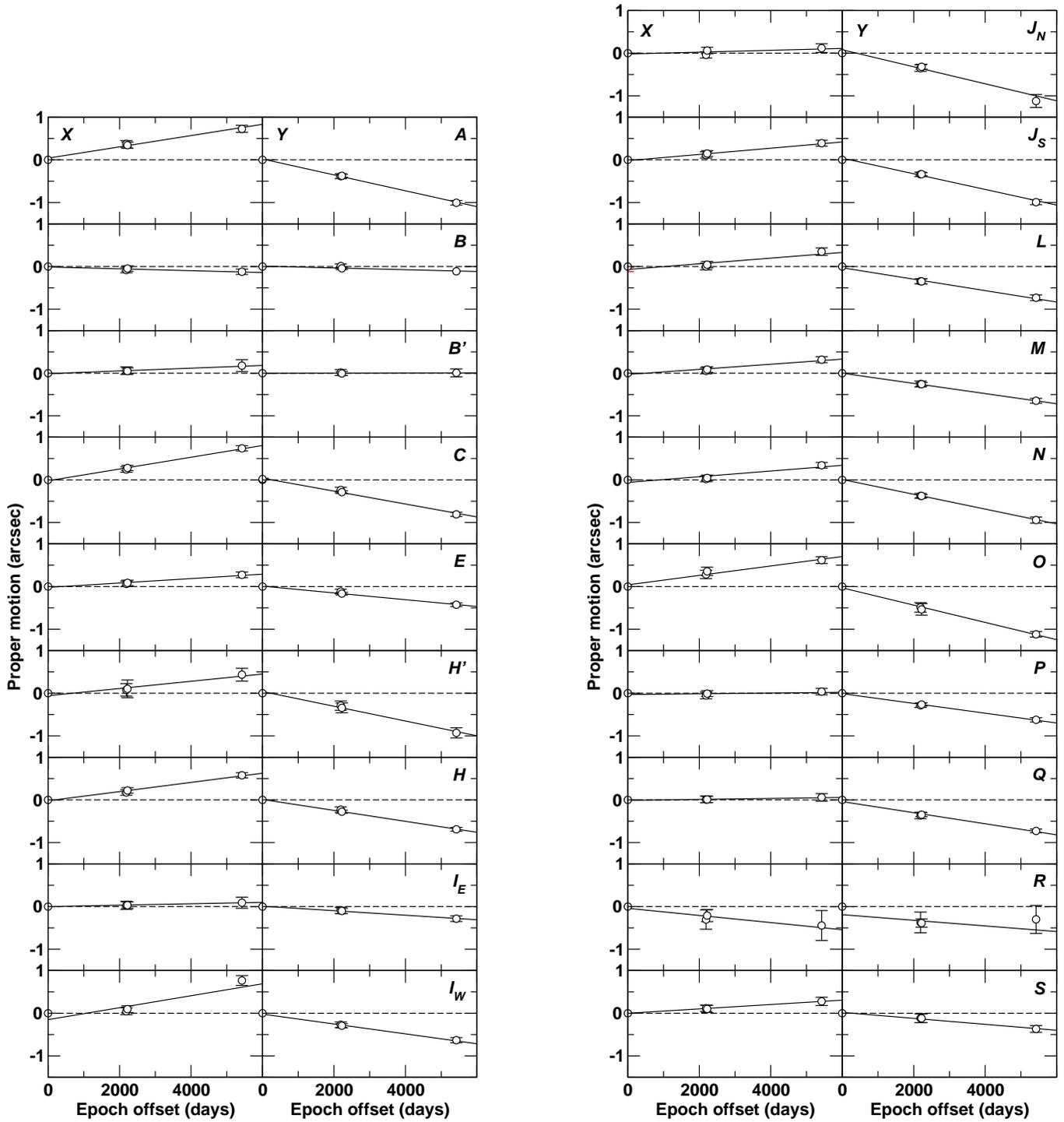

\parbox{\textwidth}{
\resizebox{0.45\textwidth}{!}{\includegraphics{1402fi2a.eps}}
\hfill
\resizebox{0.45\textwidth}{!}{\includegraphics{1402fi2b.eps}}
}
\caption{Proper motions of the HH~110 knots. 
The offsets in the $x$ direction are shown in the left panels and those in the $y$
direction in the right panels. The knot is indicated in the top right corner of
each right panel.
For each epoch,
a circle represents the displacement (in arcsec) in the
$x$-direction (left) and $y$-direction (right) of the corresponding knot,
measured from the first-epoch image (which defines the origin of the time
scale, set on 1987 Dec 18). Errors are indicated by the vertical bars. The
least squares fits derived for each knot are shown by the continuous lines.
The proper motions of the knots in the $x,y$ directions
are determined from the slopes of these lines.
\label{fig2}} 
\end{figure*}

\begin{table}
\caption[]{HH 110 Proper Motions$^a$}
\begin{tabular}{lrrrr}
\hline\hline
Knot & Offset$^b$ & $V_\mathrm{T}$ & PA & $\Delta$PA$^c$\\
     & (arcsec)   & (km~s$^{-1}$)  & (deg)&(deg)\\
\hline 
A             &   0.0 & $181.9\pm\phn5.2$ & $215\pm\phn2$ & $ 21$\\
B             &   7.9 & $ 24.2\pm\phn3.7$ & $134\pm\phn9$ & $-58$\\
B'            &  14.9 & $ 25.6\pm\phn2.6$ & $273\pm\phn6$ & $ 79$\\
C             &  22.8 & $163.0\pm\phn5.8$ & $222\pm\phn2$ & $ 28$\\
E             &  40.1 & $ 75.4\pm\phn4.0$ & $213\pm\phn3$ & $ 19$\\
H'            &  42.8 & $153.5\pm\phn7.6$ & $206\pm\phn3$ & $ 12$\\
H             &  48.4 & $133.3\pm\phn3.9$ & $220\pm\phn2$ & $ 26$\\
I$_\mathrm{E}$&  61.0 & $ 44.4\pm\phn2.0$ & $197\pm\phn1$ & $  3$\\
I$_\mathrm{W}$&  62.3 & $144.7\pm   22.3$ & $230\pm\phn7$ & $ 36$\\
J$_\mathrm{N}$&  62.4 & $159.8\pm   17.8$ & $186\pm\phn4$ & $ -8$\\
J$_\mathrm{S}$&  70.0 & $157.0\pm\phn7.3$ & $202\pm\phn2$ & $  8$\\
L             &  73.0 & $119.0\pm\phn7.7$ & $206\pm\phn6$ & $ 12$\\
M             &  81.3 & $106.6\pm\phn2.9$ & $207\pm\phn3$ & $ 13$\\
N             &  91.9 & $148.1\pm\phn4.0$ & $201\pm\phn4$ & $  7$\\
O             & 101.8 & $183.7\pm\phn5.4$ & $209\pm\phn2$ & $ 15$\\
P             & 113.6 & $ 91.5\pm\phn2.7$ & $184\pm\phn4$ & $-10$\\
Q             & 121.2 & $104.1\pm\phn5.2$ & $185\pm\phn1$ & $ -9$\\
R             & 152.0 & $ 84.9\pm   24.8$ & $128\pm   20$ & $-66$\\
S             & 179.1 & $ 68.9\pm\phn2.5$ & $216\pm\phn2$ & $ 22$\\
\hline
\end{tabular} 
\begin{list}{}{}
\item[$^a$] A distance of 460 pc has been adopted
\item[$^b$] Offset from knot A (at the 1987 epoch) along the jet outflow axis
\item[$^c$]$\Delta$PA is defined as PA$_{\rm pm}$$-$PA$_{\rm j}$
\end{list}
\label{t2}
\end{table}

In order to check the accuracy obtained for the common reference system of
the five final images, two tests were performed: 
\begin{itemize}

\item Firstly, the displacements of the reference field stars in all 
images were calculated by computing the peak of the two-dimensional
cross-correlation function between pairs of frames (i.e., the first-epoch,
1987, and each of the other images) over small boxes around the reference
stars.  The average and rms of the displacements for the four pairs of images
were less than 
$0.09  \pm 0.18$ pixels for the $x$ coordinate, and less than
$0.04 \pm 0.14$ pixels for the $y$ coordinate.

\item Secondly, to test the accuracy of the transformations described 
above, we carried out astrometry for each of the final images, 
using the $(\alpha, \delta)$ coordinates of five bright field stars 
obtained from the  USNO-B1.0 Catalogue\footnote{The USNOFS Image and 
Catalogue Archive is operated by the United States Naval Observatory, 
Flagstaff Station.}.
With this procedure we confirmed that the pixel size indeed corresponds
to $0\rlap.''35$ (see above) and that the orientation of each of
the final images differs from the true North direction by less than
$0\rlap.^{\circ}4$.

\end{itemize}
Thus, we concluded that the five final images used for the proper motion
determinations were properly scaled and aligned, and were therefore
suitable for estimating the proper motions of the HH 110 knots.

In order to calculate proper motions for the knots of HH~110, we first
defined boxes in each frame that included the emission from the individual
knots (see Table~\ref{t2} and Fig.~\ref{fig1} for the nomenclature of 
the knots and the
definition of the boxes). Then, for the four pairs of frames (i.e., the
1987 first epoch and each of the other images) we computed the
two-dimensional cross-correlation function of the emission within the
pairs of boxes, and we determined the displacement in the $x$ and $y$
coordinates through a parabolic fit to the peak of the cross-correlation
function (see Reipurth et al.\ \citealp{rei96} and L\'opez et al.\ \citealp{lop96} 
for more
details of this procedure).  For each knot, the uncertainty in the
position of the correlation peak was estimated through the scatter of the
correlation peak positions corresponding to boxes shifted from the nominal
box by 0 or $\pm2$ pixels (i.e., $0\rlap.''7$) on each of its four sides
(giving a total of $3^4$ different boxes for each knot). We adopted as
error for each coordinate  twice the rms deviation of the
positions of the cross-correlation peaks, added quadratically to the rms
alignment error.

The errors in the displacements obtained for the 1988 image were much
greater than for the rest of the images. This is probably a consequence of the
poor quality of the image. Thus, we decided to discard the data obtained
from this image from the rest of the analysis and use only the
displacements from the 1987 image to the images of 1993, 1994, and 2002.

Finally, for each knot, we performed a linear regression fit to the
displacements in each coordinate as a function of epoch offset from the
first epoch (1987). For each coordinate of each knot we fitted a straight
line to four points: a zero displacement for epoch offset zero, and the
three measured displacements. The fit was performed taking into account
the error associated with each displacement. We assigned to the zero
displacement an error equal to the quadratic average error of the three
measured displacements. The fits obtained for the $(x,y)$ coordinates of
the different knots are shown in Fig.~\ref{fig2}. From the slope of the fitted
straight line and its uncertainty we determined the proper motion
velocity ($V_\mathrm{T}$) and position angle (PA) of each knot and their
uncertainties. These values are listed in Table~\ref{t2} and displayed in 
Fig.~\ref{fig1}.
A distance to HH~110 of 460 pc has been adopted.

\subsection{Discussion}

Let us first define the HH~110 outflow axis passing through knots A to
C, which define a well aligned structure. The outflow axis defined in this
way has PA$_{\rm j}=193\rlap.^{\circ}6$. From Figure~\ref{fig1}, it is clear that, as a
general trend, the proper motion velocities show a small westward
deviation relative to this axis. This westward deviation tends to diminish
as one goes southwards from knot A. If one compares our Figure~\ref{fig1} 
with Figure 4
of Reipurth et al.\ \cite{rei96}, one sees a general agreement between the two
sets of results. However, there are substantial differences for some
of the knots, which are discussed in detail below.

At distances $>130''$ south from knot A the emitting knots show large
deviations with respect to the outflow axis, with a locus that first
curves to the E and then to the W, becoming again more or less parallel to
the outflow axis. In this region, we have measured the proper motions of
knot R (which is moving to the SE, in a direction parallel to the locus of
the jet) and condensation S (again moving parallel to the locus of the
jet, which in this position is also approximately parallel to the outflow
axis). The resulting structure is quite striking, as it apparently shows
that the knots are flowing along a curved ``channel'', in contrast to the
expected behaviour of a pure ballistic motion (note that the knot proper
motions are always parallel to the local direction of the locus of the
emitting region). However, as can be seen from Figure~\ref{fig1}, 
this conclusion
is mostly based on the SE proper motion of knot R, which has a highly
uncertain measured proper motion.  If we remove this knot from the
analysis, we would only conclude that all of the knots of HH~110 have
proper motions directed sligthly to the W of the outflow axis.

\begin{figure}
\rotatebox{-90}{\includegraphics[width=7cm]{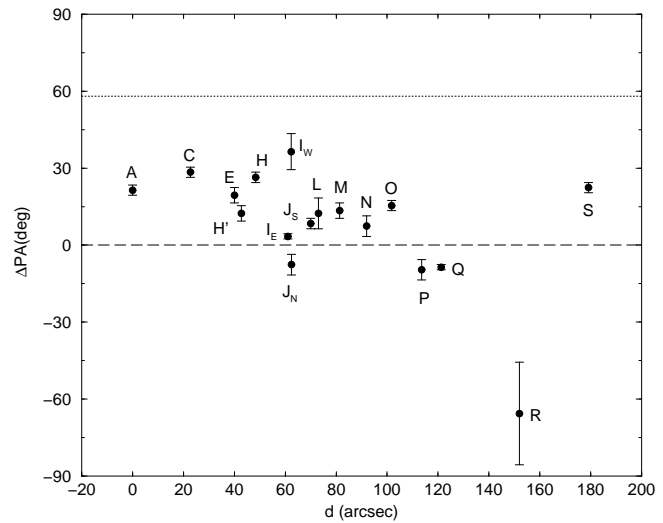}}
\caption{Deviation of the proper motion velocity direction (PA$_{\rm pm}$)  
from the axis of the HH~110 jet (PA$_{\rm j}= 193\rlap.^{\circ}6$, reference dashed line ) as a
function of distance to knot A.
The dotted line gives the
direction of the axis of HH~270 relative to the HH~110 jet axis.
 Knots B and B' are basically
stationary (see text) and have been removed from the Figure.  
\label{fig3}} 
\end{figure}

 In order to illustrate the westward shift,
we calculate the differences
$\Delta$PA=PA$_{\rm pm}-$PA$_{\rm j}$ between the directions of the proper motions
and the  direction of the HH~110 axis, PA$_{\rm j}=193^\circ.6$ (see above).
In Figure~\ref{fig3} we plot these angular differences as a function of distance
along the outflow axis. In this figure we see that knots P, Q
and J$_N$ move in a direction to the E of the HH~110 axis. 
Knot R also moves to the E (but, as we have discussed above, it is not clear
whether or not this proper motion measurement is significant), 
and the remaining 13 knots with measured proper motions move in
a direction to the W of the HH~110 axis. In this figure the
corresponding directions of the HH~110 and HH~270 jet axes
have been indicated by  dashed (for HH~110) and dotted  (for HH~270) 
lines, respectively. The directions of  
the motions of these 13 knots lie in between the HH~110
axis and the HH~270 axis (the jet that is
apparently deflected through a collision in order to form HH~110).
In addition, this plot shows that while for the knots with
distances $d<50''$ from knot A the proper motions have PA$_{\rm pm}-$PA$_{\rm j}\sim
10^\circ$ to $30^\circ$ (with an average value $\langle{\rm PA}_{pm}-{\rm
PA}_j\rangle=21^\circ$), for distances $50''<d<130''$ the proper motions have
PA$_{\rm pm}-$PA$_{\rm j}\sim -10^\circ$ to $36^\circ$ (with an average value $\langle{\rm
PA}_{pm}-{\rm PA}_j\rangle=7^\circ$). This decrease in the average angular 
offset $\langle{\rm
PA}_{pm}-{\rm PA}_j\rangle$ between the proper motions and the outflow axis
shows that there is a partial convergence between the proper motions and
the outflow axis as one moves along the HH~110 jet. However, at distances
$d>130''$ from knot A we have knots R and S (see Figure~\ref{fig3}), that again
have proper motions that are misaligned with the outflow axis. These
proper motions, as stated above, appear to be aligned with the local
direction of the curved locus of the jet.

\begin{figure}
\rotatebox{-90}{\includegraphics[width=7cm]{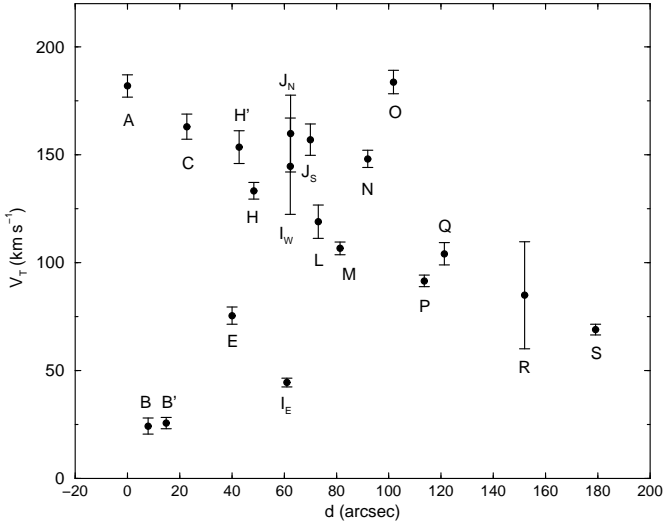}}
\caption{Moduli of the proper motion velocities of the HH~110 knots, as a
function of distance to knot A.
\label{fig4}}
\end{figure}

 In order to model the HH~270/110 system, Raga et al.\ \cite{rag02} 
performed 3-D gasdynamics simulations for a radiative jet/cloud collision. 
Two assumptions for the incident jet velocity were considered: a time-independent,
constant direction velocity ({\it Model A}) and a precessing jet of sinusoidally
varying velocity ({\it Model B}). In both models, the proper motion velocity
vectors deviate from the jet axis direction, and the predicted deviation trend 
successfully
reproduces the observed behaviour in HH~110. Thus, it might be reasonable to 
interpret the observed lack of symmetry by a jet/cloud collision.

In Figure~\ref{fig4} we show the magnitude of the proper motion velocities as a
function of distance $d$ from knot A. In the $d<90''$ region, there are 4
low velocity knots (knots B, B', E and I$_E$). All of the other knots appear
to fall on a general trend of decreasing velocities as a function of $d$
(with knot A having $V_T\simeq 180$~km~s$^{-1}$ and knot M, at
$d=81\rlap.''3$, having $V_T\simeq 110$~km~s$^{-1}$). For $d>90''$, the
proper motions of the knots have a wide range of values, from $\sim
70$~km~s$^{-1}$ for knot S up to $\sim 185$~km~s$^{-1}$ for knot O.
Note that for knot O Reipurth et al.\ \cite{rei96} 
determined a velocity of only 105~km~s$^{-1}$.

Knot B seems have remained nearly stationary  through the fifteen years elapsed
between the first-epoch and the last-epoch frames, at least
within the errors of our proper motion determinations, as shown in
Figure~\ref{fig1} and Table~\ref{t2}. In order to explore the
``stationarity'' of this region of the jet, we  marked as B' the
emission located between knot B and knot C, and we  evaluated a proper
motion for B'. The proper motion velocity found for B' is similar to the
velocity of knot B.  Thus, it appears that HH~110 has a region of a length
of $\sim 14''$ along the jet direction ( i.e., the region going from the
end of knot A to the beginning of knot C) without appreciable proper
motions between the first-epoch and last-epoch optical images. The lack of
motion of knot B is in agreement with the results of Reipurth et al.\
\cite{rei96}.

A possible explanation for the nature of the region around knot B
is that it might correspond to the location of the jet collision with the cloud.
In fact, it should be noted that the values of the [NII]/H$\alpha$ emission 
line ratio 
derived from long-slit spectra around knot B  are higher than the 
values 
found in its surroundings, around knots A and C (Reipurth et al.\ \citealp{rei96};
Riera et al.\ \citealp{rie03a}, Figure~4). In addition, 
Riera et al.\ \cite{rie03a} found  a minimum in the [SII]/H$\alpha$
line ratio around knot B  relative to its surroundings. This trend in the
[NII]/H$\alpha$ and [SII]/H$\alpha$ line ratios is
indicative of a higher gas excitation around knot B relative to its
surroundings and gives support to this interpretation, although
 alternative explanations are also possible (see next section).

Knot I is elongated in the E-W direction. It has two peaks, which we label
I$_E$ and I$_W$. The ratio of the relative intensities of these two maxima
reverses from the 1993 to the 2002 images (with I$_E$ being brighter in
1993 and I$_W$ being brighter in 2002). We find that while I$_W$ shows a
proper motion similar to the ones of other nearby knots, I$_E$ shows a
much lower proper motion (see Figure~\ref{fig4}). Therefore the proper motion of
I$_E$ appears to be somewhat doubtful, and we suspect that it might be the
result of a strong change of morphology of the emission in this region,
rather than a real motion of a physically well defined knot.

\begin{figure}
\resizebox{0.75\hsize}{!}{\includegraphics{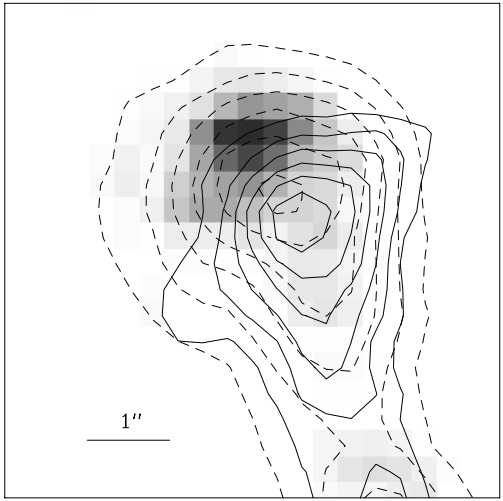}}
\caption{Close-up of HH~110 knot A from images of Dec. 1987 (grey
overlaid), Dec.1993 (dashed line contours) and Oct. 2002 (solid line
contours). North is up and East is to the left.  
\label{fig5}}
\end{figure}

 Finally, for knot A we confirm the change in morphology with time
pointed out by Reipurth et al.\  \cite{rei96}. In Figure~\ref{fig5} we show a
superposition of the 1987, 1993 and 2002 images of knot A. From this
figure we see that the morphology of this region changes quite
substantially with time. Not only does the peak of the emission move, but also
knot A develops a ``tongue'' of emitting material extending along the
outflow axis. We obtain a proper motion of $\sim 180$~km~s$^{-1}$ for knot A,
which is higher than the $\sim 50$~km~s$^{-1}$ 
determined by Reipurth et al.\ \cite{rei96}. Despite this change in the
morphology of knot A, from Figure~\ref{fig5} it is evident that the
peak of knot A shifts as a function of time. This displacement of the peak
produces the high proper motion which we find for knot A.
   
\section{Radial and full spatial velocities}

\begin{figure}
\resizebox{\hsize}{!}{\includegraphics{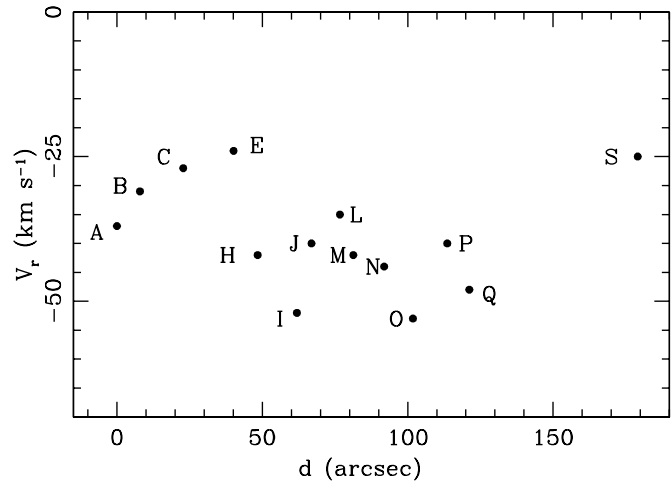}}
\caption{Radial velocities of HH~110 knots, as a function of distance to
knot A. The typical
value of the velocity error is $\sim$ 10 km s$^{-1}$ . 
\label{fig6}} 
\end{figure}

 From the long-slit spectra obtained along the HH~110 jet axis and across 
several of its brightest knots, we found that the values of 
the radial velocity 
derived from the H$\alpha$ emission  differ by less than 15--20~km~s$^{-1}$ 
(for a spectral 
resolution of $\sim 20$~km~s$^{-1}$) from the values derived 
from the [SII] emission 
(see, eg, Riera et al.\ \citealp{rie03a}, Figure 3). 
Thus, it seems reasonable  to combine  proper motions derived from [SII] 
 imaging and
radial velocities derived from the H$\alpha$ emission line 
to evaluate full spatial velocities of the knots. 
Accordingly,
we  used  the more complete HH~110 Fabry-P\'erot data of 
the H$\alpha$ emission line
from Riera et al.\ \cite{rie03b} to obtain the heliocentric 
radial velocity of the HH~110 knots.
>From these data we computed the line
profile integrated over the boxes defined for measuring the proper motions
of the knots (see Figure~\ref{fig1}), and determined the radial velocity of the
emission peak with a parabolic fit. The results are shown in Figure~\ref{fig6},
where we present the radial velocities with respect to the surrounding
cloud (which has a heliocentric radial velocity of $+23$~km~s$^{-1}$) as a
function of distance from knot A.

We can clearly identify two groups of knots in Figure~\ref{fig6}. The first group
includes knots A, B, C and E, which show an increase of the 
value of the radial velocity with distance (from $\sim-37$ km s$^{-1}$
at knot A to $\sim -$24 km s$^{-1}$ at knot E).  The second group
corresponds to the knots found at distances from $50''$ to $130''$ from knot
A (including the knots located between knot H and knot Q), which show
radial velocities ranging from $-38$ to $-53$ km s$^{-1}$. There is a
trend of more negative radial velocities at larger distances, which has
been previously reported by Riera et al.\ \cite{{rie03a},{rie03b}}.  
Finally, knot S
shows a velocity of $\sim -25$~km~s$^{-1}$ relative to the surrounding
cloud.

\begin{figure}
\resizebox{\hsize}{!}{\includegraphics[width=7cm]{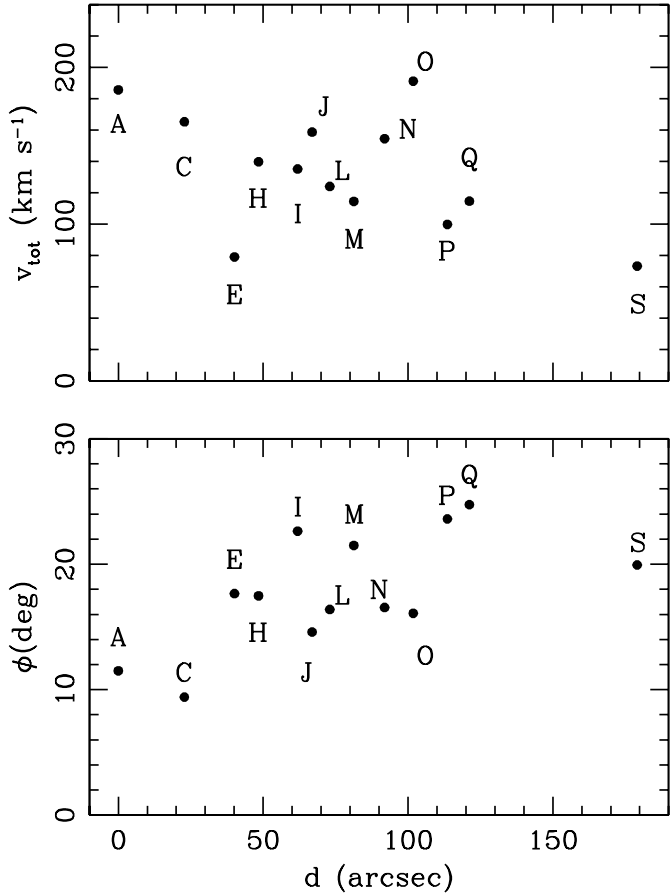}}
\caption{Total velocities (upper panel) and orientation angle, defined by
the direction of the knot motion and the plane of the sky (lower panel),
as a function of distance to knot~A. 
 The typical
value of the total velocity error is $\sim$ 12 km s$^{-1}$. The typical
value of the orientation angle error is $\sim$ 6$^\circ$.
\label{fig7}} 
\end{figure}

From the radial velocities (relative to the surrounding cloud, see above
and Figure~\ref{fig6}) and the proper motion velocities (see Table~\ref{t2}) we 
calculated the total spatial velocity $v_{\rm tot}$ of the knots, and the angle
$\phi$ between the knot motion and the plane of the sky (with positive
values of $\phi$ towards the observer). The results of this estimation are
shown in Figure~\ref{fig7}. From this figure we see that the orientation angle has
a value $\phi=9^\circ$ to $12^\circ$ for the region within $30''$ 
from knot A. At
larger values of $d$ ($>40''$), the orientation angle has values
$\phi=15^\circ$ to $25^\circ$. This range of $\sim 10^\circ$ in orientation angle
is consistent with an opening angle of $\sim 10^\circ$ (in the plane of the
sky) of this region of HH~110 (Riera et al.\ \citealp{rie03b}).

A general trend of decreasing spatial
velocity ($v_{\rm tot}$) with  distance ($d$) can be appreciated from
Figure~\ref{fig7}.
The velocities decrease from $v_{\rm tot} \simeq 190$~km~s$^{-1}$ for knot A 
to $v_{\rm tot}\simeq70$~km~s$^{-1}$ for knot S. Knots E and O are 
the ones that show the
largest deviations from the general trend of decreasing velocities as a
function of distance from knot A.

 Regarding knot B, note that the region around this knot is not fully 
stationary considering the 3-D spatial velocity, since 
we determined a continuous increase of the
radial velocity from knot A to knot C. Thus, instead of being the location of the jet
collision with the cloud, an alternative speculative
explanation might be that the jet is excavating a channel around knot B, 
with its entrance around knot A, coming out around knot C.

It should be noted that we carried out  the 
evaluation of the spatial velocity $v_{\rm tot}$ of the knots by  assuming 
that  $v_{p}$, the pattern velocity of the knots (i.e., the knot proper
motion along the jet axis, $v_{p}=V_{T}/\cos\phi$) is similar to 
the fluid velocity, $v_{f}$. This condition is fulfilled
in the case of highly supersonic, dense jets, where the knots represent
internal working surfaces (IWS). 
However, if the knots  originate from 
Kelvin-Helmholtz (K-H) instabilities, then the  ratio 
$\eta=v_{f}$/$v_{p}$ 
could 
be higher than
unity (eg, up to a factor 2, Micono et al.\ \citealp{mic98}). If 
the HH~110 knots originated from K-H instabilities instead of corresponding to IWS,  
the results we derived give a lower limit to the
full spatial velocity $v_{\rm tot}$, and the true jet motion direction $\phi$  
would 
lie closer to the plane of the sky (i.e, given the low radial
velocities found for the HH~110 knots, the values derived for $v_{\rm tot}$ 
would have to be scaled   by a factor of the order of $\eta$, while the
derived $\phi$ values would have to be scaled by a factor of the order of $\eta^{-1}$).
This $\eta$ scaling value cannot be measured from the present observations, and
has to be assumed from a model.
  
The result we found 
for the behaviour of the spatial velocity ($v_{\rm tot}$)
is very interesting, because it reverses the conclusion of
Riera et al.\ \cite{rie03a}, who deduced an increasing velocity as a function of
distance along the HH~110 flow. The origin of this difference lies in the
fact that Riera et al.\ \cite{rie03a} took the measured radial velocities and
converted them into a full spatial velocity by deprojecting these radial
velocities assuming the same value of $\phi=35^\circ$ for the 
orientation angle (with respect to
the plane of the sky) for all of the knots. The fact that we now deduce a
position-dependent orientation angle for the knots along HH~110 (see
Figure~\ref{fig7}) modifies the result of Riera et al.\ \cite{rie03a}
in such a way that the observed velocity vs. position trend is now
reversed.

The observed trend in the behaviour of the total velocity
is not predicted by the jet/cloud models of Raga et al.\ \cite{rag02}.
In particular, {\it Model A} gives velocity values significantly lower than 
the values measured in HH~110 and, in addition, predicts an increase of the 
proper motion velocity with  distance from the cloud (i.e., with the distance
from knot A). This prediction is the opposite of the observed trend 
(i.e., decreasing
proper motion velocities away from knot A).  

\section{Conclusions}

We  carried out proper motion measurements of the knots of HH~110 on a
series of five [SII]~6717+6731 CCD images covering a time baseline of
$\sim15$ years. With these measurements we obtained a better accuracy
and were able to describe the motion of more knots than in the previous
proper motion determinations of Reipurth et al.\ \cite{rei96}.
>From the Fabry-P\'erot data of Riera et al.\ \cite{rie03b}, we  obtained the
average radial velocities for the boxes around the  knots that
were used for determining the proper motions.  Combining these radial
velocities with the proper motions, we obtained the full spatial velocities
$v_{\rm tot}$ and the orientation angles $\phi$ (with respect to the plane of
the sky) of the different clumps.

We found a number of interesting effects:
\begin{itemize}

\item The proper motions of the knots in the northern part of the outflow
(i.~e., close to knot A) have an orientation angle in between the HH~110
and the HH~270 axes, westward of the HH~110 axis.

\item The knots further away from knot A  have proper motions which are
better aligned with the HH~110 axis.

\item The proper motions show a strong lack of
symmetry with respect to the outflow axis, and show a general tendency to
point towards the west. This trend 
is  predicted by the jet/cloud collision models of Raga et al.\ \cite{rag02}.

\item There is a region of $\sim 14''$ in length around knot B without 
appreciable proper motion velocity. We speculate that 
this region might trace 
the location of the jet collision with the cloud. Signatures of 
a higher gas excitation around knot B relative to its 
surroundings are found from the
[NII]/H$\alpha$ and [SII]/H$\alpha$ line ratios, which is 
in favour of this
hypothesis, although other explanations are possible.

\item Knot R, which is located in a region in which the jet has a highly
curved shape, appears to be aligned with the locus of the jet (deviating
substantially from the direction of the outflow axis). This result is
somewhat marginal, because the proper motion determined for knot R is
quite uncertain.

\item 
The spatial velocity $v_{\rm tot}$ shows a general
trend of decreasing velocities as a function of distance from knot A (i.e.,
accounting correctly for the changes in orientation
angle, one finds a ``deceleration'' of the full spatial velocity along the
HH~110 flow).  
This trend in the behaviour of the total velocity
is not predicted by the jet/cloud collision models of Raga et al.\ \cite{rag02}. 
\end{itemize}

We  conclude that HH~110 is a jet with somewhat peculiar
characteristics. 
It should be noted that
the spatial velocities of the knots along HH~110 show a decrease
by a factor of $\sim 3$ between knot A and knot S. This slowdown occurs
over a distance of $\sim 10^{18}$~cm, which is much shorter than the
distances over which one expects the braking due to interaction with the
surrounding environment to become important (Cabrit \& Raga \citealp{cab02};
Masciadri et al.\ \citealp{mas02}). 
Therefore, the strong slowdown of the spatial velocity
along HH~110 might
be an evidence for an anomalously strong interaction between the outflow and
the surrounding environment (such as might occur, e.~g., during the
deflection in a jet/cloud collision).  Alternatively, the observed
slowdown could be the result of variability in the ejection velocity, or
of changes in the jet/cloud impact region.  In order to discern between
these different possibilities, it will be necessary to carry out further
modelling of the HH~270/110 flow.

\begin{acknowledgements}

Part of this work was supported by  the Spanish MCyT grant
AYA2002-00205.  
The work of ACR was supported by the CONACyT grants 36572-E and 41320 and 
the DGAPA (UNAM) grant IN~112602.  
The NOT image was obtained using ALFOSC, which is owned by the Instituto 
de Astrof\'{\i}sica de Andaluc\'{\i}a (IAA) and operated at the NOT under 
agreement between IAA and the NBIfAFG of the Astronomical Observatory of 
Copenhagen.  
We acknowledge Gabriel G\'omez (IAC) for obtaining the October 2002 image
during the Spanish NOT Service Time.
We thank the referee for his/her valuable comments.

\end{acknowledgements}

\bibliographystyle{aa}
\bibliography{hh110}

\end{document}